\documentclass[12pt]{article}
\usepackage{graphicx} % Required for inserting images
\usepackage{amsmath,amssymb,enumerate,xcolor}
\usepackage{hyperref}

\def\slsf#1{{\slshape \sffamily #1\/}}

\title{Real-analyticity of 2-dimensional superintegrable metrics and solution of two  Bolsinov-Kozlov-Fomenko conjectures.}
\author{   Vladimir S.\ Matveev\footnote{
Institut f\"ur Mathematik, Friedrich Schiller Universit\"at Jena,
07737 Jena,  Germany  \ \ \quad {\tt  vladimir.matveev@uni-jena.de}}}

\newtheorem{theorem}{Theorem}
\newtheorem{remark}[theorem]{Remark}

\newtheorem{conjecture}[theorem]{Conjecture}

\newcommand{\const}{\textrm{const}}

\newcommand{\weg}[1]{}

 \usepackage[backend=biber]{biblatex}
\date{}

\begin{document}

\maketitle
\begin{abstract} 
We study two-dimensional Riemannian metrics which are  superintegrable in the class of polynomial in momenta  integrals. 
The study is based on our main technical result, Theorem \ref{thm:2}, which states that  the 
Poisson bracket of two polynomial in momenta integrals is an algebraic function of 
 the integrals and of the Hamiltonian. We conjecture that two-dimensional superintegrable Riemannian metrics are  necessary real-analytic in  isothermal coordinate systems, and give arguments supporting this conjecture. Small modification of the arguments, discussed in the paper, provides a  method  to construct new superintegrable systems.   We prove  a special case of the above  conjecture which is sufficient to show that  
 the metrics constructed by K. Kiyohara \cite{Kiyohara2001}, which  admit irreducible    polynomial in momenta integrals of arbitrary high degree $k$, are  not superintegrable and in particular  do  not  admit  nontrivial polynomial in momenta integral of degree less than $k$. This result    solves   Conjectures (b) and (c) explicitly formulated in \cite{BKF1995}.

\end{abstract}

\begin{flushright}
 To academician  Valery V. Kozlov \\  on the occasion of his 75th birthday
  \end{flushright}

\section{Introduction}

\label{intr1} Let $M^2$ be
a $C^\infty$ smooth connected surface  
equipped with a $C^\infty$-smooth Riemannian
metric $g=(g_{ij})$. The \slsf{geodesic flow} of the metric $g$ is the
Hamiltonian system on the cotangent bundle  $T^*M^2$ with the Hamiltonian
$H:=\tfrac{1}{2}g^{ij}p_ip_j$, where $(x,y)=(x_1, x_2)$ is a local coordinate
system on $M^2$, and $(p_x, p_y)=(p_1,p_2)$ are the correspondent
\slsf{momenta}, i.e., the dual coordinates on $T^*M^2$.

We say that a function $F:T^*M^2\to \mathbb{R}$ is an \slsf{integral} of the
geodesic flow of $g$, if $ \{F, H\}=0$, where $\{ \ , \ \}$ is the canonical
Poisson bracket on $T^*M^2$.  We say that the integral is \slsf{polynomial in
 momenta of degree $d$}, if in every local coordinate system $(x,y,p_x,p_y)$
it has the form
\begin{equation} \label{eqn:int}
\textstyle
  F(x,y,p_x,p_y)= \sum_{i=0}^da_i(x,y) p_x^{d-i}p_y^i. 
\end{equation}

For example, the  Hamiltonian $H$ itself is an integral quadratic in momenta.

Theory of two-dimensional metrics whose geodesic flows admit    polynomial in momenta  integrals is one of the oldest  parts of the theory of integrable systems,  as   nontrivial results were obtained  ar least in the 19th century. Indeed, many  classically known and studied  finite-dimensional integrable systems admit integrals which are polynomial in momenta. Moreover, if a geodesic flows  admits an integral which is analytic in momenta, then it admits an  integral which is polynomial in momenta, see e.g. \cite{bolsinov_book}. 
By \cite{KM2016},  geodesic flow of a  generic metric admits no non-trivial polynomial integral even locally. The existence of such an integral is therefore a non-trivial local differential-geometric condition on
the metric, see \cite{Kruglikov2008} for a discussion of   conditions for  the existence of  integrals of lower degrees. While locally one can prove the existence of a family of metrics, depending on $k$  functions of one variable,  admitting nontrivial  integrals polynomial in momenta of degree $k$, see e.g. \cite{BM2011, Ten}, it is not easy to construct examples on closed surfaces. It is known, see \cite{Kol1982},  that closed  surfaces of negative Euler characteristic do not admit nontrivial polynomial in momenta  integrals.  Linear and quadratic integrals on closed surfaces are   completely  understood, see e.g. \cite{BMF}.      It is still an open question whether  the geodesic flow of a  nonflat metric of the the two-torus can admit an irreducible integral of degree greater that two; Conjecture (a) of \cite{BKF1995} suggests a negative answer.   On the sphere, there exist  examples  of metrics     whose geodesic flows 
admit   polynomial in momenta  integrals  of degree $3$ and $4$ and do not admit nontrivial integrals of lower degrees, see e.g. \cite{BKF1995, DM2004, Valent2010}, and also the examples of Kiyohara \cite{Kiyohara2001} which we discuss below.

Two-dimensional metrics whose geodesic flows admit   three
functionally independent polynomial in momenta
integrals
 are called \slsf{superintegrable}.  Recall that  functions are called \slsf{functionally independent}, if their  differentials are linearly independent at almost every point. Using the methods developed in \cite{KM2016}, one can prove that if the differentials of polynomial in momenta integrals are linearly indepdent at one point of $T^*M,$ they are linearly independent at almost every point implying the functions are functionally independent. 

In our paper we study the question whether a superintegrable metric is necessary real analytic. Our ultimate  goal is the following conjecture: 

\begin{conjecture} \label{thm:1}
 Suppose a $C^\infty$-smooth metric $g= \lambda(x_1,x_2) (dx_1^2 + dx_2^2)$ on a connected 2-dimensional manifold 
  is superintegrable with 
two  polynomial in momenta  integrals  
$A= a_0(x_1, x_2)p_1^n +a_1(x_1, x_2)p_1^{n-1}p_2+\cdots  + a_n(x_1, x_2)p_2^n $,    \ $ B= b_0(x_1, x_2)p_1^k + b_1(x_1, x_2)p_1^{k-1}p_2 +  \cdots  + b_k(x_1, x_2) p_2^k$. 

Then, on  a complement to a discrete set of points,  the functions 
 $\lambda $,  $a_i$ and $b_j$ are real-analytic functions in the variables $x_1, x_2$.  
\end{conjecture}

Of course, known superintegrable, in the class of integrals polynomial in momenta,  metrics, in particular those constructed in \cite{ Koenigs, MS, valent1, valent2, duval},  support this conjecture. 

 Of course,  as two isothermal coordinate systems are connected by a holomorphic or anti-holomorphic coordinate change of the coordinate $z=x_1+ ix_2, $  real-analyticity in one isothermal coordinate system implies real-analyticity in any other.

We  suggest a method to tackle Conjecture  \ref{thm:1}, which is explained in details in \S \ref{sec:method}. The method is based on a  reduction of the problem to a very  overdetermined quasilinear  system of PDEs   with analytic coefficients.  The method is interesting besides its relation to Conjecture \ref{thm:1} as it potentially will give an algorithmic way to construct all superintegrable metrics, see \S\ref{sec:4}. In relation to   Conjecture  \ref{thm:1}, would the obtained PDE-system be of finite type, analytic dependence of solutions of ODE with real-analytic coefficients on the coefficients and on initial data will imply real-analyticity of the metric and  prove the conjecture. Unfortunately, we did not manage to show, in the  general case, that the  obtained system is of finite type, because of certain  algebraic difficulties (one needs to analyze nondegeneracy of a certain matrix  whose components come from the coefficients of the integral).  We did though managed to use the method to prove 
the following Theorem which is an important special case of Conjecture \ref{thm:1}.

\begin{theorem} \label{thm:4} Let $g $ be a two-dimensional  $C^\infty$-smooth Riemannian 
metric on the  standard disc $D= \{ (x,y)\in \mathbb{R}^2 \mid x^2 +y^2 <1\}$ whose geodesic flow admits  two polynomial in momenta integrals $A$ and $B$ such that $A$, $B$ and the Hamiltonian of the geodesic flow $H$ are functionally independent. 
Assume it has constant curvature for $x<0$.

Then, the metric is real-analytic,  in any isothermal  coordinate system, and therefore  has constant curvature on the whole disc. 
\end{theorem}

 Theorem \ref{thm:4} implies that 
  the metrics on the 2-sphere constructed by K. Kiyohara in \cite{Kiyohara2001} are not superintegrable in the class of polynomial in momenta integrals, and therefore solves Conjectures (b) and (c) explicitly stated by  A. Bolsinov,  V. Kozlov and A. Fomenko in   \cite[\S6]{BKF1995}. Note that the conjectures  are  closely related and are 
  different versions of the answer to the question  whether,  on the sphere,  there exists  a smooth metric admitting a nontrivial polynomial in momenta  integral of arbitrary large 
  degree $k$, and  admitting  no  nontrivial polynomial in momenta  integral  of a lower degree.  Let us explain that Theorem \ref{thm:4} gives a positive  answer to this question.

K. Kiyohara in \cite{Kiyohara2001}, see also discussion in   \cite[\S 3.5]{Butler}, constructed a  $C^\infty$ smooth perturbation of the standard metric on the sphere $S^2$  such that the corresponding geodesic flow admits an  integral polynomial in momenta   which we denote by  $F_K$. The perturbation depends on an arbitrary choice of a sufficiently small function of one variable with finite support and on some numerical parameters.  The numerical parameters are responsible for the degree of the integral $F_K$ which can be made arbitrary large. Kiyohara has also  shown that the integral $F_K$ is irreducible in the sense  that it can not be decomposed in the algebraic combination of integrals of lower degrees. The question whether the constructed metrics solve Conjectures (b) and (c) of \cite{BKF1995} was actively discussed in \cite{Kiyohara2001}, as   the conjectures were the main motivation for  the study. In order to solve the conjectures starting from the   example
of Kiyohara, it is necessary to show the nonexistence of an nontrivial  polynomial in momenta integral of lower degree, and this was not done. Note that Kiyohara  has shown that his metrics are Zoll metrics, in the sense that all geodesics are closed and have the same length, so their geodesic flows do admit additional integrals which are  functionally independent of $F_K$ and $H$. It is a nontrivial challenge  to show that such additional 
integrals can not be polynomial in momenta.

As mentioned above, Kiyohara's example can be viewed as perturbation  of the standard metric of the sphere. The perturbation is done on a certain open subset,  which we denote $U$, 
whose complement $S^2\setminus U$ contains an open set. Naturally, the metric of Kiyohara has constant curvature one  at $S^2\setminus U$. Kiyohara has shown that   for a generic choice of the function of one variable used in the perturbation, the metric does not have constant curvature in any nonempty open subset  of $U$.   The existence of an additional polynomial integral would  contradict   Theorem \ref{thm:4}        
and therefore  solves the above mentioned  Conjectures (b) and (c) of A. Bolsinov,  V. Kozlov and A. 
Fomenko explicitly stated in  \cite[\S 6]{BKF1995}.  It also solves \cite[Problem 3.5]{BMMT}. See also discussion in \cite[\S 3.3]{Butler} and \cite[\S 10.2]{Burns2021}.

An important step in the proof of Theorem \ref{thm:4} is interesting  by its own, and, though in this paper we will use it in dimension two only,  concerns metrics of arbitrary dimension $n$. In dimension $n$, we say that a 
metric is \slsf{ maximally superintegrable} if its geodesic flow  admits $2n-1$ functionally independent integrals which are polynomial in momenta.

\begin{theorem} \label{thm:2} Let $(M^n, g)$ be a connected $C^\infty$-smooth Riemannian manifold. Assume its  geodesic flow is maximally superintegrable and denote by $F_1=H, F_2, \dots,  F_{2n-1}$ its functionally independent integrals polynomial in momenta. 

Then, the Poisson bracket of any two of them is algebraically dependent of $F_1, \dots, F_{2n-1}$, in the sense that for every $i,j$
there exists  a polynomial $P$ of $2n$ variables which nontrivially depends on the last  variable such that 
$P( F_1,F_2,\dots, F_{2n-1}, \{F_i,F_j\})\equiv 0$.    
\end{theorem}

Note that the functional dependence of  $\{F_i ,F_j\} $ on the  functions 
$
F_1,F_2,\dots, F_{2n-1}  $ is almost trivial; the difficulty  is to show that functional dependence is in fact an algebraic one.

Note also that in dimension $2$, superintegrability is necessary maximal, as $2n-1= 4-1 =3$, so Theorem \ref{thm:2} can and will be applied in our two-dimensional  setup.

\section{ Proof of Theorem  \ref{thm:2}}

We will prove a slightly stronger result: 

\begin{theorem} \label{thm:2bis}Consider a $C^\infty$ smooth Riemannian metric $g$ on a connected $M^n$ and assume that it is superintegrable with polynomial in momenta functionally independent  
integrals $F_1,\dots, F_{2n-1}$.

Then, any polynomial in momenta integral  $F_{2n}$  is algebraically dependent of $F_1,\dots, F_{2n-1}$. 
\end{theorem}
{\bf Proof.}
 We  consider  polynomial in momenta integrals $F_1,\dots, F_{2n}$ such that first $2n-1$ of them are functionally independent. We  assume without loss of generality that each of them has   
   the same degree $\ell$. 
We may do it, since otherwise we can replace the integrals  by their  appropriate powers. 
Note that for the proof it is not important that the integral  $F_1$ coincides with the Hamiltonian $H$ of the geodesic flow and that $F_{2n}= \{F_i, F_j\}$. Of course we use the known fact that Poisson bracket of two integrals polynomial in momenta  is an integral polynomial in momenta.

Next,   consider the following linear map from 
 all  homogeneous   polynomials  $P_k$  of degree $k$  of $2n $ variables  to 
polynomial in momenta integrals  of degree $\ell k$:  
\begin{equation} \label{eq:4} P_k\mapsto P_k(F_1,...,F_{2n-1}, F_{2n})  .\end{equation}

   The dimension of the space of such polynomials is equal to 
	\begin{equation}\label{eq:2} \frac{2n(2n+1)(2n+2)...(2n+k-1)}{k!}.\end{equation} 
	By \cite[Theorem 8]{KM2016},  see also \cite{thompson,wolf},   the dimension of the space of  polynomial in momenta integrals  of degree $\ell k$ is bounded from above by 
	\begin{equation} \label{eq:3} 
	\frac{(n+\ell k - 1)!(n+\ell k)!}{(n-1)!n!(\ell k)!(\ell k +1)!}.
	\end{equation}
	Comparing (\ref{eq:2}) with (\ref{eq:3}), we see that for sufficiently large $k$, (\ref{eq:2}) is greater than (\ref{eq:3}). Indeed, \eqref{eq:3} is polynomial in $k$ of degree 
	$2 n -2$, so its natural logarithm grows as $(2  n  -2)\ln k  $ (we ignore the terms of lower order). 
	Rewriting  \eqref{eq:2} as
	$$
	\frac{2n}{k}(1+ 2 n)\left(1 + \frac{2n}{2}\right)\left(1 + \frac{2n}{3}\right)\cdots \left(1+ \frac{2n}{k-1}\right) 
	$$
	we see that its logarithm  grows  (we again ignore the terms of lower order)    as 
	$$
	\begin{array}{r}-\ln k  +   \ln\left(2n \left(1+ 2n\right) \left(1 + \frac{2n}{2}\right)\ \cdots \left(1+ \frac{2n}{k-1}\right)\right)  \sim -\ln k + 2n\left(1 +\frac{1}{2} +  \dots +\frac{1}{k-1}\right)\\ \sim    (2 n-1) \ln k . \end{array}$$
	We see that for big $k$ the dimension of the  space of  $P_k$ is greater then the dimension of   
Killing  tensors  of degree $\ell k$ implying the mapping (\ref{eq:4}) has a kernel. Therefore,  there exists a  nontrivial  polynomial $P$ such that 
	$$P(F_1,\dots ,F_{2n-1}, F_{2n})\equiv 0.$$ This polynomial must non-trivially depend on $F_{2n}$. Indeed,   the differentials of $F_i$, $i=1,\dots, 2n-1$,  are linearly independent, so if a 
    polynomial of $F_1,\dots,F_{2n-1}$ is zero then all coefficients of the polynomial are equal to zero.  Theorem \ref{thm:2bis} is proved.

\section{A system on PDEs on the coefficients of the integrals and additional equations coming from Theorem \ref{thm:2}}
\label{sec:method}
 \subsection{Setup and scheme}
We work in an isothermal coordinate  system, so that the metric has the form $g= \lambda(x_1, x_2) (dx_1^2+ dx_2^2).$
We assume that it is superintegrable with two polynomial in
momenta integrals 
\begin{eqnarray*}A & = &  a_0(x_1, x_2)p^n_1
+a_1(x_1, x_2)p^{n-1}_1 p_2 + \cdots  + a_n(x_1, x_2)p^n_2 \\ 
B & =&  b_0(x_1, x_2)p^k_1 +  \cdots  + b_{k-1}(x_1, x_2) p_1p_2^{k-1} + b_k(x_1, x_2)p^k_2.\end{eqnarray*}
We 
will construct a  quasilinear system of PDEs of fist order   whose unknowns are the coefficient $\lambda$ of the metric and the coefficients $a_i, b_i$ of the integral which is fulfilled if  the functions $A$ and $B$ are integrals.  The number of equations in this system is twice the number of unknowns; for the generic choice of the values of $a_i$ and $b_i$, the system can be solved with respect to the derivatives of unknown functions. 

The   construction goes  as follows: we start with the system of PDEs which corresponds to the properties $\{A,H\}=0$ and $\{B, H\}=0$. This system has $n+1+k+1 +1 =  n+k +3 $ 
unknowns $a_0, \dots, a_n, b_0,\dots, b_k, \lambda$, and contains  $n+2+ k+2  = n+k+4 $ equations. We see that the number of equations is not enough to express all highest (in this case, first order) derivatives. Next, we employ a trick used e.g. in Kolokoltsov \cite{Kol1982} which allows one to reduce, in one system,  the number of unknown functions by two, by the price of reducing the number of equations by two. We will then have  $n+k+2$ equations on $n+k+1$ unknowns.  
The trick is applied to a complement of the set of zeros of a certain holomorphic vector field, which is a discrete set which contribute to  the set $D$.  Next, we use Theorem \ref{thm:2} and get additional  $n+k$ equations:
they come from the condition $\{A, B\}= \Psi(H, A, B)$, where the function $\Psi$ is constructed by the polynomial $P$ in Theorem \ref{thm:2}. So we end up with   $2n+2k+2$ equations on $n+k+1 $ unknowns.  The system can be used to construct many (and hopefully, all) superintegrable metrics by an algorithmic procedure which can be realized on computer algebra software, see \S \ref{sec:4}. It will also lead to the proof of Theorem \ref{thm:4}.  

\subsection{ Employing the trick of Kolokoltsov, Darboux and Birghoff } \label{sec:trick}
The trick was known  to and was used by  classics; we explain it following V. Kolokoltsov \cite{Kol1982}. 
Take the  polynomial in momenta integral $A= a_0(x_1, x_2)p_1^n+...+a_n(x_1, x_2)p_2^n$  of
 degree $n$ for a metric $\lambda(x_1, x_2) (dx_1^2 + dx_2^2)$. 

As $\{H, A\}$ is  a homogeneous polynomial in momenta  of degree $n+1$, the   condition  $\{H, A\}=0$ is 
 equivalent to  a system of $n+2$ PDEs  on $n+2$ functions $\lambda, a_0,..., a_n$ which we view  as unknown functions.  The system is of first order  and is quasilinear, i.e., the derivatives of the unknown functions come with coefficients  which are linear expressions in the coefficients of the integrals. 

Let us now reduce the system, by an isothermal coordinate change, to a system of $n$ equations on $n$ unknown functions.

In order to do it, we pass to  the complex coordinate $z= x_1+  i x_2$, $p= \tfrac{1}{2}( p_1 - i p_2)$. In this coordinate system,  the Hamiltonian\footnote{Of course, the coefficient $\lambda$ in the Hamiltonian is still a function of $x_1, x_2$, but we will differentiate it with respect to the complex variables} is $ \frac{p \bar p}{\lambda(z, \bar z)}$ and the integral   has the form $A_0 p^n + \bar A_0 \bar p^n + A_1 p^{n-1} \bar p + \bar A_1 p \bar p^{n-1} + \dots $ with complex valued coefficients 
$A_0,..., A_{[n/2]}$, which are related to initial $a_0,...,a_n$  by some linear formulas.  Recall that   in the coordinates, the formula for the Poisson bracket  is similar to that in the real coordinates and is given, up to a constant factor,  by 
$$
\{H, F\}= \tfrac{\partial H}{\partial  p} \tfrac{\partial F}{\partial z}   + \tfrac{\partial  H}{\partial  \bar p} \tfrac{\partial F}{\partial \bar z}- \tfrac{\partial H}{\partial  z} \tfrac{\partial F}{\partial p}   - \tfrac{\partial  H}{\partial  \bar z} \tfrac{\partial F}{\partial \bar p}.  
$$

  Observe that  $\{H, F\}$ is a polynomial of degree $n+1$ in $p, \bar p$; the coefficient at  $p^{n+1}$ is $\tfrac{\partial A_0}{\partial \bar z}$. Thus,  $A_0$ is holomoprphic which in particular implies that its zeros are discrete. As a geometric object,  $A_0$ is not a function though as  it is a coefficient of a tensor field.  It is easy to check, see  e.g. \cite{Kol1982},  that, near the points such that $A_0\ne 0$, 
  under a coordinate change,   $A_0$   transforms such that     
  $\sqrt[n]{ \tfrac{1}{A_0}} dz $ is a meromorphic    differential form\footnote{Or,  equivalently, $(dz)^n/A_0$ is a meromorphic  $n$-differential}.

 After the coordinate change  
$Z_{new} = \int \sqrt[n]{ \tfrac{1}{A_0}} dz $ near points where $A_0\ne 0$,  the differential looks $dZ_{new}$, so   in the new coordinates we have  $A_0=1$, which, when we return to the initial system, means that
the coefficients $a_0,...,a_n$ satisfy the conditions $a_0 - a_2 + a_4 - \cdots =1 $
 and $a_1- a_3 + a_5 - \cdots =0$, so two  of the unknown functions, say $a_0$ and $a_1$, can be expressed as linear functions of other unknown functions. Thus, effectively we have only $n-1$  unknown coefficients of $F$, plus the unknown coefficient $\lambda$ of the metric.  As the  
 function $A_0$  is constant in the new setup, two of the equations coming  from the condition $\{A, H\}=0$ are identically fulfilled, so we effectively have 
  $n$ equations for our $n$ unknown functions.

  \begin{remark} \label{rem:2}
      The first coefficient   $A_0$ discussed above will play important role in \S \ref{sec:proof4}, let us recall its known properties.  
      \begin{itemize}
          \item  For a  homogeneous polynomial  in momenta integral   $F= P(A, B, C)$, where 
          $P$ is a polynomial of three variables with constant coefficients, and $A,B,C$ are o polynomial in momenta integrals,  the corresponding  coefficients  $F_0, A_0, B_0 $ and $C_0$ satisfy the relation $F_0= P(A_0, B_0, C_0).$
          \item For the polynomial in momenta integral  $V$  of degree $1$ the corresponding function, which we denote by $\alpha$,  determines the coefficients of the integral uniquely, as    $V(z,\bar z, p, \bar p)= \alpha p + \bar \alpha \bar p.  $
          \item For the metric $g$ of  constant positive curvature, there exist three linear in momenta integrals, $V_1, V_2 $ and $V_3$ satisfying the relation $\{V_1, V_2\}= V_3$ and $\const\, (V_1^2 + V_2^2 + V_3^2)  = H.$ Actually,  for  any metric of constant curvature  there exist three linear in momenta integrals and the Hamiltonian is a quadratic polynomial in these three linear integrals by \cite{MMS,thompson}.  Of course,  the  commutation relation and the formula for the  quadratic polynomial may  look slightly  different from that of the metric of constant positive curvature.
      \end{itemize}
  \end{remark}

  \subsection{ Getting additional  equations using Theorem \ref{thm:2}} \label{sec:additional}

We consider the 
metric $\lambda(x_1, x_2) (dx_1^2 + dx_2^2)$  and two 
polynomial in momenta integrals  $A$, of degree $n$, and $B$ of degree $k$. We assume that we already implemented the trick from \S \ref{sec:trick}, so the integral $A$ has $n-1$ unknown coefficients, say $a_3,a_4,\dots, a_{n+1}$.

By Theorem \ref{thm:2},  the integral $F:= \{A, B\}$ is algebraically dependent of $H, A, B$, that is there exists a nontrivial  polynomial $P$ of one variable with coefficients which are polynomials $C_i(H,A,B)$, with constant coefficients,   in  three variables $H, A$ and $ B$,  which vanishes on $F$: 
\begin{equation}\label{eq:pol}
    P(F):= C_\ell(H, A, B) F^\ell+   C_{\ell-1}(H, A, B) F^{\ell-1}+\cdots+C_0(H, A, B)=0.
\end{equation}
Take a point $(x,p)\in T^*M^2$ such that $p\ne 0$.  Let us show that without loss of generality we may assume that for a certain discrete set $D$ of points, at any point  $x\in M^2\setminus D$, for almost 
all $p\in T_x^*M^2$, $F$ is simple root of $P$.   

Assume contrary, so locally we have infinitely many points  $x_i$ such that at every $p\in T_{x_i}^*M^2$ the function 
 $F(x_i, p)$ is a multiple root of multiplicity $m$ of  $P$. Consider the function  
$P^{(m-1)}(F):=  \tfrac{d^{m-1}}{dt^{m-1}} P(t)_{|t=F} $. It is an algebraic expression of integrals of our geodesic flow and is therefore an  integral. It  vanished at all points $x_i$. Now, as explained in \cite[\S 2.2]{KM2016}, if a polynomial in momenta 
integral vanishes at  sufficiently many points, it vanished identically implying $F$ has multiplicity $m$ at all points. 

Further, we assume that $F$   is a simple  root of $P$ at all points we are working in.  Then, by the implicit function theorem,  there exists (in a neighborhood of almost every point $(x,p)$) 
a (real-analytic) function $\Psi$ of three arguments such that $\Psi(H,A,B)= F= \{A, B\}$.  This gives us additional equations
\begin{equation} \label{eq:additional}
\{A, B\}= \Psi(H,A,B).
\end{equation}
We would like to emphasize, that the left hand side is quasilinear expression of the first order 
in the coefficients of $A$ and $B$, and the   right hand side depends on the coefficients of $A, B$, on $\lambda$, and on certain constants which are coefficients of the polynomials $C_i$ from \eqref{eq:pol}, but does not depend on the derivatives of the coefficients of  $A, B$ and of $\lambda$.
As the left hand side is a polynomial in momenta of degree $n+k-1$, the additional equation \eqref{eq:additional} gives us $n+k$ equations. 
Our  count   of equations and unknowns is summarized in the following table:

\begin{tabular}{c|c|c c}
 & \# equations & \# unknowns     & explanation \\ \hline
$\{H,A\}=0$ &  $n$  &  $n$  &  coefficients of $A$ and $\lambda$ \\ 
 $\{H,B\}=0$ & $k+2$ & $k+1$ &  coefficients of $B$  \\  
$\{A,B\}=\Psi$ &  $n+k$ &  $0$ &  no uncounted  unknowns  \\
\hline
together:  &  $2(n+k +1)$  & $n + k + 1$& \\ \hline 
\end{tabular}

Let us now 
view the whole system of $2(n+k +1)$ PDEs as a linear algebraic system on the first derivatives of unknown function. 
It is a system on  $2(n+k+1)$  equations on $2(n+k+1)$ unknown first derivatives, so the coefficient matrix  is a square matrix depending on $x\in M^2$.

If  for some values of the coefficients of $A$, $B$ and of $\lambda$
its kernel is trivial at a point $(\hat x_1, \hat x_2), $ then it is trivial in a small neighborhood of $(x_1, x_2)$. Then, locally, there exists at most one 
 solution with these initial data, and it is real-analytic. Indeed,   for an arbitrary point $(x_1, x_2)$ which is close to $(\hat x_1, \hat x_2)$ we conisder the segment 
 $t \mapsto (t x_1 + (1-t) \hat x_1, t x_2 + (1-t) \hat x_2).$ Our system of PDEs   gives a systems of ODEs of the Euler type for the restrictions of the unknowns to the segment. The uniqueness of the solution follows then from the standard results on the  uniqueness   of the solutions of ODEs. Next, as the system of PDEs is real-analytic, the system of ODEs   is also real-analytic and the standard results on the dependence of solution on initial data and on coefficients imply that the solution of the system of PDEs, if exists, is real-analytic as well.

Let us now consider the case when the kernel is not trivial. Let us consider the prolongation of the equation, i.e., derive  the equation with respect to $x_1$ and $x_2$.  As the system was quasilinear, the new  system is  also quasilinear; moreover, the coefficients of the 2nd order terms   contain no derivatives. The number of equations is now doubled, and the number of unknown functions  is multiplied by $1.5, $ as for any unknown coefficient we had  three second derivatives and  two first derivatives, so the number of highest derivatives went up with factor $1.5$. 

We again view this system of PDE as the linear system on the second derivatives of unknown functions; it has now more equations than unknowns.  If the kernel of the corresponding matrix is  trivial, then by the same argument we see that the  solution, if exists, is real-analytic. 
We can continue this procedure, derive the equations one more time, consider it as a linear system on higher derivatives and conclude that  if in the case we  can solve the system with respect to the highest derivatives,  the solution is necessary real-analytic. 

Our computer experiments indicate that for  certain prolongation of the system is always solvable with respect to higherst derivatives. As mentioned in the introduction, we did not manage to prove the results for all degrees of the integrals.

\subsection{ A method which possibly lead to computer-algebra realizable construction of (all?)
superintegrable geodesic flows} \label{sec:4}

In \S \ref{sec:additional} we explained how to obtain a system  $$\{H,A\}=0,\  
 \{H,B\}=0, \  
\{A,B\}=\Psi(H,A, B)$$   of $2k+2n +2$ equations on $k+n+1$ unknown functions.  Solutions $(\lambda, a_i, b_i)$ of this system  correspond to two-dimensional  metrics with  superintegrable geodesic flows and to the corresponding integrals. The freedom in constructing of the system, besides  the choice of degrees $k, n$,  is   the choice of the algebraic function $\Psi$, which is essentially the same as the choice of polynomial  $P$ of 4 variables $H, A, B , \{A, B\}$. The prolongation (i.e., differentiation with respect of $x_1,x_2$) of the system contains the same information as the prolongation  
 of the system   $\{H,A\}=0$, $\{H,B\}=0$, 
$P(H, A, B, \{A, B\})  =0$, which is linear in second derivatives of the unknown functions. Therefore, compatibility conditions for this systems can be calculated algorithmically. They  are rational  relations, with coefficients coming from the coefficients of $P$,  on the unknown coefficients $a_i, b_i, \lambda$ and their first derivatives. Such relations can in theory  be resolved using algorithmic computer algebra methods, e.g., the Gröbner basis method, which will possibly  lead to a description in quadratures of all    superintegrable geodesic flows. 
We plan to attack this problem using this circle of ideas   elsewhere.

\section{Proof of Theorem \ref{thm:4}} \label{sec:proof4}

  We assume that the metric is polynomially superintegrable with integrals $A$ and $B$.  The coordinates we will work in will always assumed to be isothermal so 
  $g=  {\lambda}(x_1, x_2) (dx_1^2 + dx_2^2)$. 
  
  It is sufficient to show that if a point  $X$ satisfied the condition that the cotangent space to this point has points in which the differentials of $H, A $ and $B$ are linearly independent and    lies in the closure of an open connected 
set $U\subset D$  such that at every point of  $U$ the metric has constant curvature, the   metric is real-analytic near the point $X$ and therefore has constant curvature in a neighborhood of $X$. Indeed, this would imply that the closure of the set of points having a neighborhood in which the metric has constant curvature is an open subset of $D$; since it is tautologically a closed subset, it must coincide with $D$ and we are done. Next, we may assume without loss of generality 
that the cotangent plane to the point $X$ contains a point where the differentials of   $H, A$ and $B$ are linearly independent, we may do it since as explained in \S \ref{sec:additional} the points where this property does not hold form a discrete set, so the complement is still connected and the open-closed argument above still works. 

In order to show the above statement,  we first recall, see also  Remark \ref{rem:2},  that metrics of constant curvature have precisely  three functionally independent polynomial in momenta integrals of degree $1$, which we denote by $V_1, V_2, V_3$. This statement is trivial as linear in momenta integrals are essentially the same as Killing vector fields and metrics of constant curvature have three independent  Killing vector fields.  We note also that for any point $X$ and for almost every point of the cotangent plane to the point the differentials of $V_1, V_2$ and $V_3$ are linearly independent. 

Next, recall that 
any polynomial in momenta integral $A$ is an algebraic combination of linear integrals:  \begin{equation} \label{eq:equation} A= P_A(V_1, V_2, V_3),  \ B= P_B(V_1, V_2, V_3), H= P_H(V_1, V_2 , V_3)\end{equation}
with some polynomials $P_A, P_B, P_H$ of three variables with constant coefficients.  This result was proved e.g. in \cite{thompson, MMS}. 

In what follows we will work in a small neighborhood of the point $(X, P)$ of the cotangent plane to $X$ such that at this point the differentials of the integrals $A,B$, $H$ are linearly independent and the differentials of the integrals $V_1, V_2$, $V_3$ are also linearly independent. In what follows we will reduce the problem to a system of PDEs by taking the Poisson bracket of the integrals.  As a restriction of a polynomial to an open set determines the polynomial, restricting to a small neithborhood of the point $ (X, P)$  does not loose any relevant for the proof information.

Locally, by the implicit function Theorem,  there exist analytic functions $\Phi_1, \Phi_2, \Phi_3$ such that (for  almost all points $(x,p)\in T^*U$) 
\begin{equation} \label{eq:Phi} V_1= \Phi_1(H, A, B), \ V_2= \Phi_2(H, A, B), \  V_3= \Phi_3(H, A, B). \end{equation}
Indeed,  the differentials of the functions $V_1, V_2, V_3$ and of the functions $H,A, B$ are connected by the Jacobi $3 \times 3$-matrix of the polynomial mapping $$(V_1, V_2, V_3)\stackrel{\eqref{eq:equation}}{\mapsto} (H, A, B) .$$  Since the differentials of  $V_1, V_2, V_3$ and of   $H,A, B$  
are linearly independent, the Jacobi matrix   is nondegenerate. 

 Next, consider  the 
functions $V_1, V_2, V_3$ 
given by \eqref{eq:Phi} in a small  neighborhood of the point $(X,P)$: at the points  lying over 
$U$, they are linear in momenta. They are well-defined functions in a small neighborhood of  $(X,P)$.   Of course,  we do not know a priory whether they are linear in momenta at the point which do not lie over $U$. 

Moreover, the following  analog of the relation $\{A, B\}= \Psi(H,A,B) $
 still holds for the integrals $V_1, V_2, V_3$ and has the form $$\{V_1, V_2\}= \textrm{linear combination, with constant coefficients,  of $V_1, V_2, V_3$} .$$ Indeed, 
 $$\{ \Phi_1(H,A,B), \Phi_2(H, A, B)   \}= \Psi(H,A, B) \left(
 \tfrac{\partial \Phi_1}{\partial A}\tfrac{\partial \Phi_2}{\partial B}   -  \tfrac{\partial \Phi_2}{\partial A}\tfrac{\partial \Phi_1}{\partial B} \right). 
 $$
 Here,   $\Psi$  denotes the   real-analytic function  such that $\{A, B\}= \Psi(A,B,H)$. The existence of such function follows from   Theorem \ref{thm:2}, see discussion in \S \ref{sec:additional}. 
 We see that $\{V_1, V_2\}$ depends on $H, A, B$ real analytically. Therefore, we have  that the equality 
 $$\{\Phi_1 , \Phi_2   \} = \textrm{linear combination, with constant coefficients,   of $\Phi_1 , \Phi_2 , \Phi_3$}$$ 
   holds over $U$, and therefore   everywhere.

   In what follows, we will additionally assume that our metric $g$ has positive (constant)  curvature for $x<0$. This is sufficient for the  solution of  Conjectures (b) and (c) of \cite{BKF1995},
   as   Kiyohara's example is a perturbation of the standard metric which has positive curvature.  The proof for zero  and negative curvature is completely analogous. Indeed, though  the commutations relation \eqref{eq:com}, and also the third formula in \eqref{eq:equation1}  may  look slightly differently for the flat  metric, 
   it does not really affect the proof.   
   
   As $g$ has constant positive  curvature on a certain open nonempty  subset, 
   without loss of generality, we may think \begin{equation}\label{eq:com}\{V_1, V_2\}=V_3.\end{equation}

We now consider the system of equations \begin{equation}\label{eq:quasilinear}
    \{H,A\}=0 ,  \  
 \{H,B\}=0 ,  \ 
 \{A,B\}=\Psi(A, B, H). 
\end{equation}We view the system  as a system of partial differential 
equations on the unknown coefficients $a_i, b_i$ of the integrals and on the 
coefficient $\lambda$ of the metric.  
Let us now differentiate, sufficiently many times,  the system with respect to variables $x_1,x_2$. We would like to show that one can solve the obtained system with respect to the highest derivatives of the unknown functions. Note that the system  \eqref{eq:quasilinear} is quasilinear, which implies that  the  coefficients near highest derivatives of the unknown functions are linear expressions in functions  $a_i, b_i,   \lambda$.  

Assume the degrees of $A$ and $B$ are $n$ and $k$, respectively,  and assume   $k\le n$. 
It is well-known, see e.g. \cite{KM2016,thompson},  that one can solve the $n$th derivatives of the equation 
$ \{H,A\}=0 $ with respect to  the $n+1$st derivatives  of the unknown coefficients $a_i$ of $A$. The solution linearly depends  on the $n+1$st  derivatives of $\lambda$, with coefficients which depend on 
$ a_i $ and $\lambda$,  and  the free term which is an explicit algebraic 
expression  in the  lower derivatives of $a_i$ and of $\lambda$. 

Similarly, as $k\le n$, we can solve the $n$th derivatives of the equation 
$ \{H,B\}=0 $ with respect to  the $n+1$st derivatives  of the unknown coefficients $b_i$ of $B$. 

Next,  let us show that all  the $n+1$st derivatives of the function $\lambda$ can  be obtained as real-analytic functions of  lower derivatives of the unknown functions $a_i, b_i, \lambda$. 
In fact, we will see that one  can obtain the first derivatives of  $\lambda$ as real-analytic functions of   the unknown function and of the coordinates. 

It is more convenient to work in the complex coordinates $z= x_1+ i x_2$, $\bar z= x_1- i x_2.$  In these coordinates, the metric has the form $\lambda(z, \bar z) dz d\bar z$. 
Next, we denote by $\alpha_i$ the holomorphic function corresponding to the linear in momenta integral $V_i$, see Remark \ref{rem:2}. We  will assume without loss of generality that  $\alpha_1=1$. Indeed, we can do it by a coordinate change as explained in \S \ref{sec:trick}. 
Let us show that the functions $\alpha_2$ and $\alpha_3$ can be constructed by the holomorphic functions $A_0$ and $B_0$. 

As recalled  in Remark \ref{rem:2}, we have  in view of \eqref{eq:equation} the system 
\begin{equation} \label{eq:equation1}
   \begin{array}{l} A_0= P_A(\alpha_1, \alpha_2, \alpha_3),  \\  B_0= P_B(\alpha_1, \alpha_2, \alpha_3), \\0= P_H(\alpha_1, \alpha_2 , \alpha_3)= {(\alpha_1^2 + \alpha_2^2 + \alpha_3^2)\const.  } \end{array}
\end{equation}
The last equation, in view of the assumption $\alpha_1=1,$ means  $\alpha_2= \sqrt{1-\alpha_3^2}. $ We may assume without loss of generality that the linear  integrals $V_i$ are not proportional at the point $X$, so  $\sqrt{1-\alpha_3^2}$ is a well-defined holomorphic function. As explained above, for one and therefore for almost every values of $\alpha_1, \alpha_2, \alpha_3$, the Jacobi matrix of the polynomial mapping given by \eqref{eq:equation1} is nondegenerate.   
As the function $\alpha_3$ is not constant, we may assume without loss of generality that the derivative of the function   $ P_A(1, \sqrt{1-\alpha_3^2}, \alpha_3)$  with respect to $\alpha_3$ is not zero. Indeed, we can achieve this by  replacing $X$ by a  point lying close to  $X$, but still lying on the boundary between regions  where the curvature is constant and where it is not constant.  Then, by the implicit  function theorem, we can solve the equation
$A_0=  P_A(1, \sqrt{1-\alpha_3^2}, \alpha_3)$ with respect to $\alpha_3, $ the solution depends analytically on $A_0$. This will also gives us the function $\alpha_2= \sqrt{1-\alpha_3^2}$.

Next, consider the equation $\{ V_1, H\}=0$,  $\{V_2, H\}=0$ and $\{ V_3, H\}=0$.
As $\alpha_1$ and $\alpha_2$ are already holomorphic, each of these equations  is  essentially one equation. 
They read as follows: 
\begin{equation}\label{eq:eq2}
\begin{array}{rlc}   \frac{\partial\lambda}{\partial z}   +  \frac{\partial\lambda}{\partial \bar z}& =& 0, \\
\lambda \frac{\partial\alpha_2  }{\partial z}   + \lambda \frac{\partial\bar \alpha_2  }{\partial \bar z}  +\alpha_2  \frac{\partial\lambda}{\partial z}   + \bar \alpha_2  \frac{\partial\lambda}{\partial \bar z} &=&0, \\ 
\lambda \frac{\partial\alpha_3  }{\partial z}   + \lambda \frac{\partial\bar \alpha_3  }{\partial \bar z}  +\alpha_3  \frac{\partial\lambda}{\partial z}   + \bar \alpha_3  \frac{\partial\lambda}{\partial \bar z} &=&0. \\ 
\end{array}\end{equation}
The functions $\alpha_i$ there are analytic functions constructed by $A_0$ and $B_0$.  We view the functions  $A_0$ and $B_0$  as certain ``given'' functions  and not as  part of unknown functions; so $\alpha_i$  and their derivatives 
are also ``known".  They are holomorphic in isothermal coordinate, and therefore analytic.
Clearly, one can express $\frac{\partial\lambda}{\partial z}  $ and $\frac{\partial\lambda}{\partial \bar z}  $ from  the equation \eqref{eq:eq2}.  
Finally, we obtain that in a neighborhood of $X$,   the $n+1$st derivatives of $\lambda, a_i, b_i$ are expressed in lower derivatives and in analytic functions $A_0$  and  $B_0$ via real-analytic formulas. As explained in \S\ref{sec:trick}, this implies that the metric is real-analytic in a neighborhood  of $X$ and we are done.

  \subsection*{Acknowledgement. } I thank the DFG (projects 455806247 and 529233771), and the ARC  (Discovery Programme DP210100951) for their support, S. Scapucci for useful discussion, and the anonymous referee for useful suggestions.

\printbibliography
\end{document}